\documentclass[12pt]{article}
\usepackage{amsmath}
\usepackage{amssymb}
\usepackage{amsfonts}
\usepackage{graphics}

\numberwithin{equation}{section}

\title{
Quasi-exactly solvable extended trigonometric P\"oschl-Teller potentials with position-dependent mass}
\author{C. Quesne\\
{\small \sl Physique Nucl\'eaire Th\'eorique et Physique Math\'ematique,}\\ 
{\small \sl Universit\'e Libre de Bruxelles, Campus de la Plaine CP229,} \\ 
{\small \sl Boulevard~du Triomphe, B-1050 Brussels, Belgium} \\
{\small \sl cquesne@ulb.ac.be}}
\date{ }
\begin{document}
\baselineskip=22pt plus 1pt minus 1pt
%%%%%%%%%%%%%%%%%%%%%%%%%%%%%%%%%%%%%%%%%%%%%%%%%%%%%%%%%%
\maketitle

\begin{abstract} 
Infinite families of quasi-exactly solvable position-dependent mass Schr\"odinger equations with known ground and first excited states are constructed in a deformed supersymmetric background. The starting points consist in one- and two-parameter trigonometric P\"oschl-Teller potentials endowed with a deformed shape invariance property and, therefore, exactly solvable. Some extensions of them are considered with the same position-dependent mass and dealt with by a generating function method. The latter enables to construct the first two superpotentials of a deformed supersymmetric hierarchy, as well as the first two partner potentials and the first two eigenstates of the first potential from some generating function $W_+(x)$ [and its accompanying function $W_-(x)$]. The generalized trigonometric P\"oschl-Teller potentials so obtained are thought to have interesting applications in molecular and solid state physics.
\end{abstract}

\noindent
Short title: Extended trigonometric P\"oschl-Teller potentials

\noindent
PACS Nos.: 03.65.Fd, 03.65.Ge
%
%========================================================================
%
\newpage
\section{Introduction}

The trigonometric P\"oschl-Teller (TPT) potential (also called P\"oschl-Teller I or Darboux-P\"oschl-Teller potential) is one of the most valuable exactly solvable (ES) potentials in nonrelativistic quantum mechanics \cite{poschl, flugge}. It is indeed close to potentials widely used in molecular physics to describe out-of-plane bending vibrations and in solid state physics to provide models for one-dimensional crystals \cite{antoine}. It is also related to the Scarf I potential \cite{scarf} via simple changes of variable and of parameters \cite{cq12}.\par
%
%-----------------------------------------------------------------------------------------------------------------
%
The TPT potential is (translationally) shape invariant (SI) in supersymmetric (SUSY) quantum mechanics \cite{genden}. Such a property provides an easy way of solving the corresponding Schr\"odinger equation \cite{cooper}. First- and second-order SUSY transformations have been used to generate new potentials whose spectrum slightly differs from the TPT one \cite{contreras}. Recently, some extensions of the TPT potential have been extensively studied (see, {\it e.g.}, \cite{cq08, odake09, odake11, gomez14, bagchi15, grandati} and references quoted therein) in connection with the new concepts of exceptional orthogonal polynomials \cite{gomez09}, para-Jacobi polynomials \cite{calogero}, or confluent Darboux transformations \cite{fernandez}.\par
%
%--------------------------------------------------------------------------------------------------------------
%
On the other hand, considering a position-dependent mass (PDM) instead of a constant one in the Schr\"odinger equation is known to play an important role in many physical problems, such as the study of electronic properties of semiconductor heterostructures \cite{bastard, weisbuch}, quantum wells and quantum dots \cite{serra, harrison}, helium clusters \cite{barranco}, graded crystals \cite{geller}, quantum liquids \cite{arias}, metal clusters \cite{puente}, nuclei \cite{ring, bonatsos}, nanowire structures \cite{willatzen}, and neutron stars \cite{chamel}.\par
%
%-------------------------------------------------------------------------------------------------------------
%
Exact solutions of PDM Schr\"odinger equations may provide a conceptual understanding of some physical phenomena, as well as a testing ground for some approximation schemes. Such solutions may belong not only to ES Schr\"odinger equations, for which all the eigenstates can be found explicitly by algebraic means, but also to quasi-exactly solvable (QES) equations, for which only a finite number of eigenstates can be derived in this way for some ad hoc couplings, while the remaining ones can only be obtained through numerical calculations.\par
%
%--------------------------------------------------------------------------------------------------------
%
The generation of PDM and potential pairs leading to such exact solutions has been achieved by various methods (see, {\it e.g.}, \cite{cq06} and references quoted therein). In particular, on taking advantage of the known equivalence of PDM problems to those arising from a deformation of the canonical commutation relations \cite{cq04}, it has been shown that several well-known ES potentials in a constant mass background remain ES for a well chosen PDM \cite{bagchi05, cq09}. This has been achieved by using a deformed supersymmetric (DSUSY) approach and a deformed shape invariance (DSI) concept. Among those potentials, one finds both the one- and two-parameter TPT potentials.\par
%
%--------------------------------------------------------------------------------------------------
%
The aim of the present work is to construct infinite families of QES extensions of these ES PDM and TPT potential pairs with known ground and first excited states. For such a purpose, we plan to use a recently devised generating function method \cite{cq18} (see also \cite{voznyak}), generalizing a procedure known for constant mass problems \cite{tkachuk}.\par
%
%--------------------------------------------------------------------------------------------------------------
%
This paper is organized as follows. In sec.~2, the description of PDM Schr\"odinger equations in DSUSY and the DSI property are reviewed, then the corresponding results for the ES one- and two-parameter TPT potentials are recalled. In sec.~3, the generating function method for constructing PDM Schr\"odinger equations with known ground and first excited states is presented. Such a procedure is then applied to extensions of one- and two-parameter TPT potentials in secs.~4 and 5, respectively. Finally, sec.~6 contains the conclusion.\par
%
%=======================================================================
%
\section{Deformed supersymmetric approach to the trigonometric P\"oschl-Teller potentials with position-dependent mass}

The standard Schr\"odinger equation
\begin{equation}
  \left(\hat{p}^2 + V(x) - E\right) \psi(x) = 0,  \label{eq:SE}
\end{equation}
where $\hat{p} = - {\rm i} d/dx$ and $\hbar = 1$, is known to be ES for the one- and two-parameter TPT potentials \cite{poschl, flugge, cooper}, defined by
\begin{equation}
  V(x) = A(A-1) \sec^2 x, \qquad - \tfrac{\pi}{2} < x < \tfrac{\pi}{2}, \qquad A>1,  \label{eq:TPT-1}
\end{equation}
and
\begin{equation}
  V(x) = A(A-1) \sec^2 x + B(B-1) \csc^2 x, \qquad 0 < x < \tfrac{\pi}{2}, \qquad A, B>1,  \label{eq:TPT-2}
\end{equation}
respectively.\par
%
%------------------------------------------------------------------------------------------------------------
%
Let us replace $\hat{p}$ by $\hat{\pi} = - {\rm i} \sqrt{f(x)} (d/dx) \sqrt{f(x)}$, where $f(x)$ is some positive and smooth parameter-dependent function and $\hat{\pi}$ is assumed to be Hermitian with respect to the measure $dx$ \cite{cq04}. Then the standard commutation relation $[\hat{x}, \hat{p}] = {\rm i}$ is changed into $[\hat{x}, \hat{\pi}] = {\rm i} f(x)$ and the conventional Schr\"odinger equation (\ref{eq:SE}) becomes
\begin{align}
  (\hat{H} - E) \psi(x) &= \left(\hat{\pi}^2 + V(x) - E\right) \psi(x) \nonumber \\ 
  &= \left(- \sqrt{f(x)} \frac{d}{dx} f(x) \frac{d}{dx} \sqrt{f(x)} + V(x) - E\right) \psi(x) = 0. \label{eq:def-SE}
\end{align}
This deformed Schr\"odinger equation can be interpreted as a PDM one,
\begin{equation}
  \left(- m^{-1/4}(x) \frac{d}{dx} m^{-1/2}(x) \frac{d}{dx} m^{-1/4}(x) + V(x) - E\right) \psi(x) = 0,
  \label{eq:PDM-SE}
\end{equation}
where $m(x) = 1/f^2(x)$. As is well known, the noncommutativity of $m(x)$ with the differential operator $d/dx$ creates an ordering ambiguity in PDM Schr\"odinger equations \cite{vonroos}. The ordering obtained in (\ref{eq:PDM-SE}) is that chosen by Mustafa and Mazharimousavi \cite{mustafa}, from which other orderings can be taken care of by replacing $V(x)$ by some effective potential $V_{\rm eff}(x)$ including derivatives of $m(x)$.\par
%
%------------------------------------------------------------------------------------------------------
%
Bound state wavefunctions $\psi_n(x)$ of eq.~(\ref{eq:def-SE}) (or, equivalently, (\ref{eq:PDM-SE})) have to be square integrable on the interval of definition $(x_1, x_2)$ of $V(x)$ with respect to the measure $dx$ and, in addition, must ensure the Hermiticity of $\hat{H}$ or, equivalently, that of $\hat{\pi}$, imposing that \cite{bagchi05}
\begin{equation}
  |\psi_n(x)|^2 f(x) = \frac{|\psi_n(x)|^2}{\sqrt{m(x)}} \to 0 \qquad \text{for $x \to x_1$ and $x \to x_2$.}
  \label{eq:Hermiticity}
\end{equation}
\par
%
%--------------------------------------------------------------------------------------------------------------
%
A DSUSY approach to eq.~(\ref{eq:def-SE}) consists in considering a pair of partner Hamiltonians, defined on the same interval $(x_1, x_2)$,
\begin{equation}
  \hat{H}_{1,2} = \hat{\pi}^2 + V_{1,2}(x) + E_0, \qquad V_{1,2}(x) = W^2(x) \mp f(x) \frac{dW}{dx},
  \label{eq:V1,2}
\end{equation}
where $E_0$ denotes the ground state energy of eq.~(\ref{eq:def-SE}) and $V_1(x)$ is the rescaled potential $V_1(x) = V(x) - E_0$ \cite{cq04, bagchi05}. The superpotential $W(x)$ in eq.~(\ref{eq:V1,2}) can be expressed in terms of the ground state wavefunction $\psi_0(x)$ of $\hat{H}_1$ through
\begin{equation}
  W(x) = - f(x) \frac{d}{dx} \log \psi_0(x) - \frac{1}{2} \frac{df}{dx}
\end{equation}
or, conversely,
\begin{equation}
  \psi_0(x) \propto f^{-1/2} \exp\left(- \int^x \frac{W(x')}{f(x')} dx'\right).  \label{eq:psi0}
\end{equation}
\par
%
%------------------------------------------------------------------------------------------------
%
The two first-order differential operators
\begin{equation}
  \hat{A}^{\pm} = \mp \sqrt{f(x)} \frac{d}{dx} \sqrt{f(x)} + W(x),  \label{eq:A}
\end{equation}
allow to rewrite the two partner Hamiltonians (\ref{eq:V1,2}) as
\begin{equation}
  \hat{H}_1 = \hat{A}^+ \hat{A}^- + E_0, \qquad  \hat{H}_2 = \hat{A}^- \hat{A}^+ + E_0,
\end{equation}
so that the latter intertwine with $\hat{A}^+$ and $\hat{A}^-$ as $\hat{A}^- \hat{H}_1 = \hat{H}_2 \hat{A}^-$ and $\hat{A}^+ \hat{H}_2 = \hat{H}_1 \hat{A}^+$. The ground state wavefunction $\psi_0(x)$ of $\hat{H}_1$ is annihilated by the operator $\hat{A}^-$, while the ground state wavefunction $\psi'_0(x)$ of $\hat{H}_2$ is transformed by $\hat{A}^+$ into the first excited state wavefunction $\psi_1(x)$ of $\hat{H}_1$.\par
%
%--------------------------------------------------------------------------------------------------------
% 
This procedure can in principle be iterated by considering $\hat{H}_2$ as a new starting Hamiltonian, thereby obtaining another DSUSY pair of partner Hamiltonians
\begin{equation}
  \hat{H}'_{1,2} = \hat{\pi}^2 + V'_{1,2}(x) + E'_0, \qquad V'_{1,2}(x) = W^{\prime 2}(x) \mp f(x) 
  \frac{dW'}{dx},  \label{eq:V'1,2}
\end{equation}
where
\begin{equation}
  V'_1(x) + E'_0 = V_2(x) + E_0.  \label{eq:V'-V}
\end{equation}
Then the first excited state wavefunction of $\hat{H}_1$ with energy $E_1 = E'_0$,
\begin{equation}
  \psi_1(x) \propto \hat{A}^+ \psi'_0(x),  \label{eq:psi1}
\end{equation}
can be obtained from the ground state wavefunction of $\hat{H}'_1 = \hat{H}_2$, given by
\begin{equation}
  \psi'_0(x) \propto f^{-1/2} \exp\left(- \int^x \frac{W'(x')}{f(x')} dx'\right).  \label{eq:psi'0}
\end{equation}
\par
%
%------------------------------------------------------------------------------------------------------------
% 
Equation (\ref{eq:V'-V}) can be rewritten as
\begin{equation}
  W^2(x) + f(x) \frac{dW}{dx} = W^{\prime 2}(x) - f(x) \frac{dW'}{dx} + E_1 - E_0  \label{eq:W-W'}
\end{equation}
in terms of the two superpotentials $W(x)$ and $W'(x)$. Such a condition can be satisfied, in particular, whenever, up to some additive constant $R$, $V_1(x)$ and $V_2(x)$ are similar in shape and differ only in the parameters that appear in them. In such a case, $V(x)$ is said to be deformed shape invariant (DSI) and eq.~(\ref{eq:W-W'}) is referred to as the DSI condition. The latter can then be generalized to any neighbouring members of a DSUSY hierarchy and the whole bound state spectrum of eq.~(\ref{eq:def-SE}) can be easily derived.\par
%
%-------------------------------------------------------------------------------------------------------------
%
{}For the one-parameter TPT potential (\ref{eq:TPT-1}), the DSI condition is satisfied for the deforming function
\begin{equation}
  f(x) = 1 + \alpha \sin^2 x, \qquad -1 < \alpha \ne 0,  \label{eq:f-1}
\end{equation}
corresponding to a PDM $m(x) = (1 + \alpha \sin^2 x)^{-2}$, and for the two superpotentials \cite{bagchi05}
\begin{align}
  W(x) &= \lambda \tan x, \qquad \lambda = \tfrac{1}{2}(1 + \alpha + \Delta), \qquad \Delta = \sqrt{(1+
      \alpha)^2 + 4A(A-1)}, \\ 
  W'(x) &= \lambda' \tan x, \qquad \lambda' = \lambda + 1 + \alpha.
\end{align}
The first two partner potentials read
\begin{align}
  V_1(x) &= A(A-1) \sec^2 x - A(A-1) - \tfrac{1}{2}(1 + \alpha + \Delta), \\
  V_2(x) &= [A(A-1) + (1+\alpha) (1+\alpha+\Delta)] \sec^2 x - A(A-1) \nonumber \\
  & \quad {}- \tfrac{1}{2}(1+2\alpha) (1+\alpha+\Delta),
\end{align}
while the ground and first excited state energies of $V(x)$ are given by
\begin{align}
  E_0 &= \lambda(\lambda-\alpha) = A(A-1) + \tfrac{1}{2}(1+\alpha+\Delta), \\
  E_1 &= (\lambda+1)^2 - \alpha(\lambda-1) = A(A-1) + \tfrac{1}{2}(5+5\alpha+3\Delta).
\end{align}
More generally, the whole bound state spectrum is obtained as
\begin{align}
  E_n &= (\lambda+n)^2 - \alpha(\lambda-n^2) \nonumber \\
  &= A(A-1) + \tfrac{1}{2}(1+\alpha+\Delta) + (1+\alpha+\Delta)n + (1+\alpha)n^2, \nonumber \\
  & \quad n=0, 1, 2, \ldots, 
\end{align}
with the corresponding wavefunctions
\begin{equation}
  \psi_n(x) \propto f^{- \frac{1}{2}\left(\frac{\lambda}{1+\alpha} + 1\right)} (\cos x)^{\frac{\lambda}{1+
     \alpha}} C_n^{\left(\frac{\lambda}{1+\alpha}\right)}(t), \qquad t = \sqrt{\frac{1+\alpha}{f}} \sin x,
\end{equation}
expressed in terms of Gegenbauer polynomials \cite{cq09}. Note that, in this case, eq.~(\ref{eq:Hermiticity}) does not provide any additional condition since it is automatically fulfilled for square integrable functions $\psi_n(x)$ on $(-\pi/2, \pi/2)$.\par
%
%-------------------------------------------------------------------------------------------------------------
%
{}For the two-parameter TPT potential (\ref{eq:TPT-2}), the deforming function, obtained in \cite{bagchi05},\footnote{The results given in \cite{bagchi05, cq09} are for the Scarf I potential. They have been transformed here for the TPT potential by using the changes of variable and parameters given in \cite{cq12}.} writes
\begin{equation}
  f(x) = 1 + \alpha \cos 2x, \qquad 0 < |\alpha| < 1,  \label{eq:f-2}
\end{equation}
with corresponding PDM $m(x) = (1+\alpha \cos 2x)^{-2}$, and the two superpotentials are
\begin{align}
  W(x) &= \lambda \tan x - \mu \cot x, \qquad \lambda = \tfrac{1}{2}(1-\alpha+\Delta_1), \qquad
      \mu = \tfrac{1}{2}(1+\alpha+\Delta_2), \nonumber \\
  \Delta_1 &= \sqrt{(1-\alpha)^2 + 4A(A-1)}, \qquad \Delta_2 = \sqrt{(1+\alpha)^2 + 4B(B-1)}, \\
  W'(x) &= \lambda' \tan x - \mu' \cot x, \qquad \lambda' = \lambda+1-\alpha, \qquad \mu' = \mu+1+\alpha.
\end{align}
The first two partner potentials read
\begin{align}
  V_1(x) &= A(A-1) \sec^2 x + B(B-1) \csc^2 x - A(A-1) - B(B-1) - \tfrac{3}{2}(1-\alpha^2) \nonumber \\ 
  & \quad{}- (1+\alpha)\Delta_1 - (1-\alpha)\Delta_2 - \tfrac{1}{2}\Delta_1\Delta_2, \\
  V_2(x) &= [A(A-1) + (1-\alpha)(1-\alpha+\Delta_1)] \sec^2 x \nonumber \\
  & \quad{} + [B(B-1) + (1+\alpha)(1+\alpha+\Delta_2] \csc^2 x - A(A-1) - B(B-1) \nonumber \\
  & \quad {} - \tfrac{1}{2}(3+5\alpha^2) - (1-\alpha)\Delta_1 - (1+\alpha)\Delta_2 - \tfrac{1}{2}
      \Delta_1 \Delta_2,
\end{align}
while the ground and first excited state energies of $V(x)$ are given by
\begin{align}
  E_0 &= (\lambda+\mu)^2 + 2\alpha(\lambda-\mu) = A(A-1) + B(B-1) + \tfrac{3}{2}(1-\alpha^2) 
     \nonumber \\
  & \quad{} + (1+\alpha)\Delta_1 + (1-\alpha)\Delta_2 + \tfrac{1}{2}\Delta_1\Delta_2, \\
  E_1 &= (\lambda+\mu+2)^2 + 6\alpha(\lambda-\mu) - 4\alpha^2 = A(A-1) + B(B-1) \nonumber \\
  & \quad{} + \tfrac{19}{2}(1-\alpha^2) + 3(1+\alpha)\Delta_1 + 3(1-\alpha)\Delta_2 + \tfrac{1}{2}\Delta_1
      \Delta_2.
\end{align} 
The whole bound state spectrum is obtained as 
\begin{align}
  E_n &= (\lambda+\mu+2n)^2 + 2\alpha(\lambda-\mu)(2n+1) - 4\alpha^2 n^2 \nonumber \\
  &= A(A-1) + B(B-1) + \tfrac{3}{2}(1-\alpha^2) + (1+\alpha)\Delta_1 + (1-\alpha)\Delta_2 + \tfrac{1}{2}
      \Delta_1 \Delta_2 \nonumber \\
  & \quad{} + 2n[2(1-\alpha^2) + (1+\alpha)\Delta_1 + (1-\alpha)\Delta_2] + 4(1-\alpha^2)n^2, \nonumber\\    
  &\qquad n=0, 1, 2, \ldots,
\end{align}
with the corresponding wavefunctions \cite{cq09}
\begin{align}
  \psi_n(x) &\propto f^{-\frac{1}{2}\left(1 + \frac{\lambda}{1-\alpha} + \frac{\mu}{1+\alpha}\right)}
      (\cos x)^{\frac{\lambda}{1-\alpha}} (\sin x)^{\frac{\mu}{1+\alpha}} P_n^{\left(\frac{\mu}{1+\alpha} -
      \frac{1}{2}, \frac{\lambda}{1-\alpha} - \frac{1}{2}\right)}(t), \nonumber \\
  t &= \frac{\cos 2x + \alpha}{1 + \alpha\cos 2x},
\end{align}
expressed in terms of Jacobi polynomials.\par
%
%===================================================================
%
\section{Generating function method for PDM Schr\"odinger equations with two known eigenstates}
\setcounter{equation}{0}

Let us start from eq.~(\ref{eq:W-W'}) relating the superpotentials $W(x)$ and $W'(x)$ of the first two steps of a DSUSY hierarchy and  let us define the two functions \cite{cq18, voznyak}
\begin{equation}
  W_+(x) = W'(x) + W(x), \qquad W_-(x) = W'(x) - W(x).  \label{eq:W_+-W_-}
\end{equation}
In terms of the latter, eq.~(\ref{eq:W-W'}) can be rewritten as
\begin{equation}
  f(x) \frac{dW_+}{dx} = W_+(x) W_-(x) + E_1 - E_0.
\end{equation}
Hence, $W_-(x)$ can be expressed in terms of $W_+(x)$ and the energy difference $E_1-E_0$ as
\begin{equation}
  W_-(x) = \frac{f(x) dW_+(x)/dx + E_0 - E_1}{W_+(x)}.  \label{eq:W_-}
\end{equation}
\par
%
%--------------------------------------------------------------------------------------------------------
%
The generating function method starts from two functions $W_+(x)$ and $W_-(x)$ that are compatible, {\it i.e.}, such that there exists some positive constant $E_1-E_0$ satisfying eq.~(\ref{eq:W_-}). The two superpotentials $W(x)$ and $W'(x)$ are then obtained from eq.~(\ref{eq:W_+-W_-}) as $W(x) = \frac{1}{2}(W_+ - W_-)$ and $W'(x) = \frac{1}{2}(W_+ + W_-)$. The starting potential $V(x)$ and its ground state energy $E_0$ are determined from $W(x)$ through eq.~(\ref{eq:V1,2}) and the ground state wavefunction is derived from eq.~(\ref{eq:psi0}). The knowledge of $E_1-E_0$ and $E_0$ provides the first excited state energy $E_1$, while a combination of eqs.~(\ref{eq:A}), (\ref{eq:psi1}), (\ref{eq:psi'0}), and (\ref{eq:W_+-W_-}) leads to the corresponding wavefunction
\begin{align}
  \psi_1(x) &\propto \left(- f \frac{d}{dx} - \frac{1}{2} \frac{df}{dx} + W\right) f^{-1/2} \exp\left(-
     \int^x \frac{W'(x')}{f(x')} dx'\right) \nonumber \\
  &\propto [W'(x) + W(x)] f^{-1/2} \exp\left(- \int^x \frac{W'(x')}{f(x')} dx'\right) \nonumber \\
  &\propto W_+(x) f^{-1/2} \exp\left(- \int^x \frac{W'(x')}{f(x')} dx'\right). \label{eq:psi1-W+}
\end{align}
The construction of the first two bound state wavefunctions $\psi_0(x)$ and $\psi_1(x)$ of $V(x)$ is of course only valid provided such functions satisfy both the square integrability condition on $(x_1,x_2)$ and the additional restriction (\ref{eq:Hermiticity}). As observed in sec.~2, the latter is automatically fulfilled for the deforming functions $f(x)$ considered for the one- and two-parameter TPT potentials provided the wavefunctions are square integrable.\par
%
%======================================================================
%
\section{Extensions of the one-parameter trigonometric P\"oschl-Teller potentials}
\setcounter{equation}{0}

In the present section, we will deal with an infinite family of extensions of the one-parameter TPT potential (\ref{eq:TPT-1}), defined by
\begin{equation}
  V^{(m)}(x) = \sum_{k=1}^{2m+1} A_{2k} \sec^{2k}x, \qquad - \frac{\pi}{2} < x < \frac{\pi}{2}, \qquad
  m = 1, 2, \ldots,  \label{eq:Vm}
\end{equation}
with $A_{4m+2}>0$. As it is obvious, the $m=0$ case would give back potential (\ref{eq:TPT-1}) with $A_2 = A(A-1)$. Our aim consists in showing that parameters $A_2$, $A_4$, \ldots, $A_{4m}$ can be found in terms of $A_{4m+2}$ and $\alpha$ in such a way that the PDM Schr\"odinger equation (\ref{eq:def-SE}) with $f(x)$ and $V(x)$ given by (\ref{eq:f-1}) and (\ref{eq:Vm}), respectively, has known ground and first excited states. The corresponding PDM will then be
\begin{equation}
  m(x) = (1 + \alpha \sin^2x)^{-2}, \qquad -1 < \alpha \ne 0.
\end{equation}
\par
%
%-------------------------------------------------------------------------------------------------------
%
\begin{figure}
\begin{center}
\includegraphics{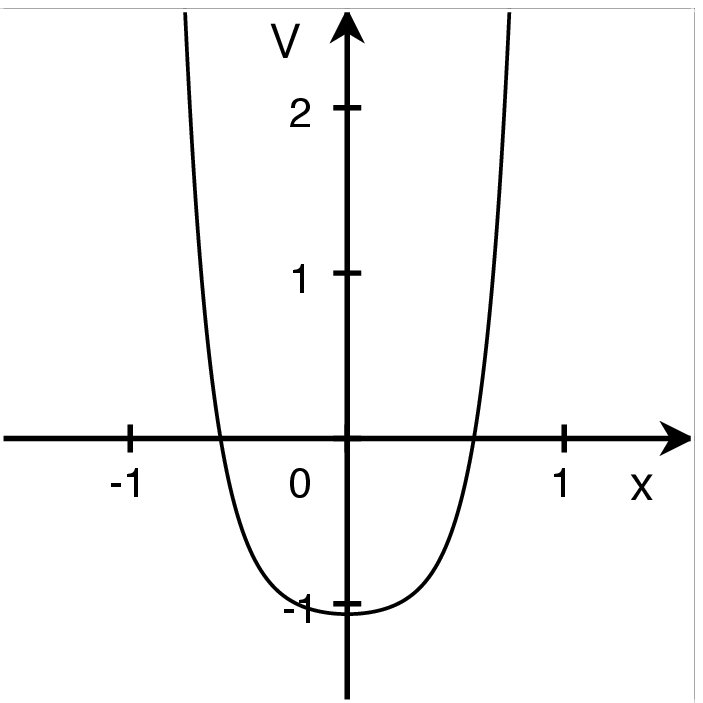}
\caption{Plot of the  extended potential $V^{(1)}(x)$ with $A_6=1$ and $\alpha=-1/2$. The ground and first excited state energies are $E_0 = 19/16$ and $E_1 = 115/16$. The PDM reads $m(x) = \left(1 - \frac{1}{2} \sin^2x\right)^{-2}$.}
\end{center}
\end{figure}
\par
%
%-------------------------------------------------------------------------------------------------------------
%
\begin{figure}
\begin{center}
\includegraphics{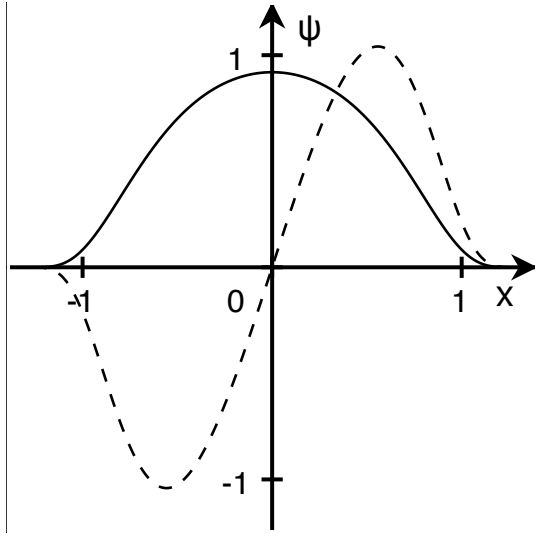}
\caption{Plots of the ground state wavefunction $\psi_0(x)$ (solid line) and of the first excited state wavefunction $\psi_1(x)$ (dashed line) for the potential displayed in fig.~1.}
\end{center}
\end{figure}
\par
%
%-----------------------------------------------------------------------------------------------------------------
%
{}For such a purpose, let us consider the generating functions
\begin{align}
  W_+(x) &= 2\sqrt{A_{4m+2}} \sum_{k=0}^m \frac{(2m+1)!!}{(2k+1)!! (2m-2k)!!} (1+\alpha)^{k-m}
      (\tan x)^{2k+1}, \label{eq:W_+1}\\
  W_-(x) &= (2m+1) (1+\alpha) \tan x.  \label{eq:W_-1}
\end{align}
To prove their compatibility, we have to show that there exists some positive constant $E_1-E_0$ satisfying eq.~(\ref{eq:W_-}). Straightforward calculations show that
\begin{align}
  \frac{dW_+}{dx} &= 2\sqrt{A_{4m+2}} \Biggl\{\frac{(2m+1)!!}{(2m)!!} (1+\alpha)^{-m} \nonumber \\
  &\quad{} + \sum_{k=1}^m \frac{(2m+1)!!}{(2k-1)!! (2m-2k)!!} (1+\alpha)^{k-m} (\tan x)^{2k}\Biggr\}  
       \sec^2 x,
\end{align} 
\begin{align}
  f \frac{dW_+}{dx} &= 2\sqrt{A_{4m+2}} \Biggl\{\frac{(2m+1)!!}{(2m)!!} (1+\alpha)^{-m} \nonumber \\
  &\quad{} + (2m+1)\sum_{k=0}^m \frac{(2m+1)!!}{(2k+1)!! (2m-2k)!!} (1+\alpha)^{k+1-m} 
       (\tan x)^{2k+2}\Biggr\},
\end{align}
and
\begin{align}
  & W_+ W_-  \nonumber \\
  &= 2\sqrt{A_{4m+2}} (2m+1)\sum_{k=0}^m \frac{(2m+1)!!}{(2k+1)!! (2m-2k)!!} 
  (1+\alpha)^{k+1-m} (\tan x)^{2k+2},
\end{align}
so that we indeed get
\begin{equation}
  E_1-E_0 = 2\sqrt{A_{4m+2}} \frac{(2m+1)!!}{(2m)!!} (1+\alpha)^{-m} > 0.  \label{eq:diff-1}
\end{equation}
\par
%
%----------------------------------------------------------------------------------------------------
%
{}From (\ref{eq:W_+1}) and (\ref{eq:W_-1}), we then obtain the two superpotentials $W(x)$ and $W'(x)$ in the form
\begin{equation}
  W(x) = \sum_{k=0}^m \lambda_k (\tan x)^{2k+1}, \qquad W'(x) = \sum_{k=0}^m \lambda'_k 
  (\tan x)^{2k+1},  \label{eq:W-1}
\end{equation}
where
\begin{equation}
\begin{split}
  \lambda_0 &= \sqrt{A_{4m+2}} \frac{(2m+1)!!}{(2m)!!} (1+\alpha)^{-m} - \frac{1}{2}(2m+1)(1+\alpha), \\
  \lambda'_0 &= \lambda_0 + (2m+1)(1+\alpha), \\
  \lambda_k &= \lambda'_k = \sqrt{A_{4m+2}} \frac{(2m+1)!!}{(2k+1)!! (2m-2k)!!} (1+\alpha)^{k-m}, \qquad
     k=1, 2, \ldots, m.
\end{split}. \label{eq:lambda-1}
\end{equation}
\par
%
%----------------------------------------------------------------------------------------------------------------
%
To determine $V(x)$ and $E_0$ from eqs.~(\ref{eq:V1,2}), (\ref{eq:W-1}), and (\ref{eq:lambda-1}), it is convenient to proceed in two steps: first to express $V_1(x)$ as an expansion in $\tan^2 x$, 
\begin{equation}
  V_1(x) = \sum_{k=0}^{2m+1} a_k (\tan x)^{2k},  \label{eq:a-1}
\end{equation}
then to reexpress it as an expansion in $\sec^2 x$ by making use of the relation $\tan^2 x = \sec^2 x -1$. In such a way, we obtain $E_0$ and the parameters $A_{2k}$, $k=1, 2, \ldots, 2m$ of eq.~(\ref{eq:Vm}) as
\begin{equation}
  E_0 = \sum_{k=0}^{2m+1} (-1)^{k+1} a_k, \qquad A_{2k} = \sum_{l=k}^{2m+1} (-1)^{l-k} \binom{l}{k}
  a_l.
\end{equation}
From the values of the coefficients $a_k$ in eq.~(\ref{eq:a-1}), we get
\begin{align}
  E_0 &= \frac{1}{4} (2m+1) (1+\alpha) [2m+1 + (2m+3)\alpha] \nonumber \\
  & \quad{} + \sqrt{A_{4m+2}} \frac{(2m+1)!!}{(2m)!!} (1+\alpha)^{-m} \nonumber \\
  & \quad{} \times \biggl[1 + 2(2m+1) \sum_{k=1}^{m+1} (-1)^k 
      \frac{(2m)!!}{(2k-1)!! (2m-2k+2)!!} (1+\alpha)^k\biggr] \nonumber \\
  & \quad{} - A_{4m+2} (1+\alpha)^{-2m} \biggl[\sum_{k=1}^{m+1} (-1)^k S^{(m,k)}_{0,k-1} (1+
      \alpha)^{k-1} \nonumber \\
  & \quad{} + \sum_{k=m+2}^{2m+1} (-1)^k S^{(m,k)}_{k-m-1,m} (1+\alpha)^{k-1}\biggr],  \label{eq:E0-1}
\end{align}
\begin{align}
  A_2 &= \frac{1}{4}(2m+1)(2m+3) (1+\alpha)^2 \nonumber \\
  & \quad{} + 2(2m+1) \sqrt{A_{4m+2}} \sum_{k=1}^{m+1} (-1)^k k \frac{(2m+1)!!}{(2k-1)!! (2m-2k+2)!!} 
      (1+\alpha)^{k-m} \nonumber \\
  & \quad{} - A_{4m+2} (1+\alpha)^{-2m} \biggl[\sum_{k=1}^{m+1} (-1)^k k S^{(m,k)}_{0,k-1}
      (1+\alpha)^{k-1} \nonumber \\
  & \quad{} +\sum_{k=m+2}^{2m+1} (-1)^k k S^{(m,k)}_{k-m-1,m} (1+\alpha)^{k-1}\biggr],
\end{align}
\begin{align}
  A_{2k} &= - 2(2m+1) \sqrt{A_{4m+2}} \sum_{l=k}^{m+1} (-1)^{l-k} \binom{l}{k} 
      \frac{(2m+1)!!}{(2l-1)!! (2m-2l+2)!!} (1+\alpha)^{l-m} \nonumber \\
  & \quad{} + A_{4m+2} (1+\alpha)^{-2m} \biggl[\sum_{l=k}^{m+1} (-1)^{l-k} \binom{l}{k} S^{(m,l)}_{0,l-1}
      (1+\alpha)^{l-1} \nonumber \\
  & \quad{} + \sum_{l=m+2}^{2m+1} (-1)^{l-k} \binom{l}{k} S^{(m,l)}_{l-m-1,m} (1+\alpha)^{l-1} \biggr], 
      \qquad 2 \le k \le m+1,
\end{align}
\begin{equation}
  A_{2k} = A_{4m+2} \sum_{l=k}^{2m+1} (-1)^{l-k} \binom{l}{k} S^{(m,l)}_{l-m-1,m} (1+\alpha)^{l-2m-1},
  \qquad m+2 \le k \le 2m,
\end{equation}
where we have introduced finite sums $S^{(m,k)}_{a,b}$, $0 \le a \le b \le m$, defined by
\begin{equation}
  S^{(m,k)}_{a,b} = \sum_{l=a}^b \frac{[(2m+1)!!]^2}{(2l+1)!! (2k-2l-1)!! (2m-2l)!! (2m-2k+2l+2)!!}.
\end{equation}
Equations (\ref{eq:diff-1}) and (\ref{eq:E0-1}) yield the first excited state energy $E_1$.\par
%
%-------------------------------------------------------------------------------------------------------
%
It remains to determine the wavefunctions $\psi_0(x)$ and $\psi_1(x)$ from eqs.~(\ref{eq:psi0}) and (\ref{eq:psi1-W+}), respectively. For such a purpose, it is useful to express the ratios $W(x)/f(x)$ and $W'(x)/f(x)$ in terms of a new variable $y=\cos x$. For the former, for instance, we get the relation
\begin{align}
  \frac{W(x)}{f(x)} &= \sin x \bigl[y^{2m+1} (1+\alpha-\alpha y^2)\bigr]^{-1} \sum_{k=0}^m \lambda_k
       y^{2m-2k} (1-y^2)^k \nonumber \\
  &= \sin x \bigl[y^{2m+1} (1+\alpha-\alpha y^2)\bigr]^{-1} \sum_{p=0}^m \biggl[\sum_{l=0}^p (-1)^l
       \binom{m+l-p}{l} \lambda_{m+l-p}\biggr] y^{2p} \nonumber \\
  &= \sin x \biggl(\sum_{\kappa=0}^m \frac{C_{2\kappa+1}}{y^{2\kappa+1}} 
       + \frac{\alpha C_1 y}{1+\alpha-\alpha y^2}\biggr),
\end{align}
with 
\begin{align}
  C_{2\kappa+1} &= \frac{1}{(1+\alpha)^{m+1-\kappa}} \sum_{p=0}^{m-\kappa} \alpha^{m-\kappa-p}
       (1+\alpha)^p \sum_{l=0}^p (-1)^l \binom{l+m-p}{l} \lambda_{l+m-p}, \nonumber \\
  & \qquad \kappa=0, 1, \ldots, m,
\end{align}
from which the integration in eq.~(\ref{eq:psi0}) is straightforward. The results read
\begin{equation}
  \psi_0(x) \propto f^{-\frac{1}{2}(C_1+1)} (\cos x)^{C_1} \exp\left(-\sum_{\kappa=1}^m 
  \frac{C_{2\kappa+1}}{2\kappa} \sec^{2\kappa}x\right)
\end{equation}
and
\begin{align}
  \psi_1(x) &\propto f^{-\frac{1}{2}(C_1+2m+2)} (\cos x)^{C_1} \exp\left(-\sum_{\kappa=1}^m 
     \frac{C_{2\kappa+1}}{2\kappa} \sec^{2\kappa}x\right) \nonumber \\
  & \quad{}\times \sum_{k=0}^m \biggl\{\biggl[\sum_{l=0}^k (-1)^{k-l} \binom{m-l}{k-l} 
     \frac{(2m+1)!!}{(2l+1)!! (2m-2l)!!} (1+\alpha)^{l-m}\biggr] \nonumber \\
  & \quad{}\times \sin^{2k+1}x\biggr\}.
\end{align}
The function $f(x)$ having a finite value $1+\alpha$ for $x \to \pm \pi/2$, the behaviour of $\psi_0(x)$ and $\psi_1(x)$ at the interval boundaries is determined by that of $\exp\left[- C_{2m+1} \sec^{2m}x/(2m)\right]$,
where $C_{2m+1}= \sqrt{A_{4m+2}}/(1+\alpha) > 0$, thereby showing that such functions are square integrable, as it should be.\par
%
%------------------------------------------------------------------------------------------------
%
As an illustration, let us present some detailed results for the $m=1$ case. The potential reads
\begin{align}
  V^{(1)}(x) &= \left[\frac{15}{4} (1+\alpha)^2 + 3(1+4\alpha) \sqrt{A_6} 
     + \frac{3(4\alpha^2-1)}{4(1+\alpha)^2} A_6\right] \sec^2 x \nonumber \\
  & \quad{} - 3\sqrt{A_6} \left[2(1+\alpha) + \frac{\alpha}{1+\alpha} \sqrt{A_6}\right] \sec^4 x
     + A_6 \sec^6 x.  \label{eq:V(1)}
\end{align}
Its ground and first excited state energies are
\begin{align}
  E_0 &= \frac{3}{4}(1+\alpha)(3+5\alpha) + \frac{3(-1+2\alpha+4\alpha^2)}{2(1+\alpha)} \sqrt{A_6}
      + \frac{(1-2\alpha)^2}{4(1+\alpha)^2} A_6, \\
  E_1 &= \frac{3}{4}(1+\alpha)(3+5\alpha) + \frac{3(1+2\alpha+4\alpha^2)}{2(1+\alpha)} \sqrt{A_6}
      + \frac{(1-2\alpha)^2}{4(1+\alpha)^2} A_6,      
\end{align}
with corresponding wavefunctions
\begin{align}
  \psi_0(x) &\propto f^{\frac{1}{4} - \frac{\sqrt{A_6}}{4(1+\alpha)^2}} 
      (\cos x)^{\frac{\sqrt{A_6}}{2(1+\alpha)^2} - \frac{3}{2}} \exp\left(- \frac{\sqrt{A_6}}{2(1+\alpha)}
      \sec^2 x\right),  \label{eq:V(1)gs} \\
  \psi_1(x) &\propto f^{-\frac{5}{4} - \frac{\sqrt{A_6}}{4(1+\alpha)^2}} 
      (\cos x)^{\frac{\sqrt{A_6}}{2(1+\alpha)^2} - \frac{3}{2}} \sin x [3 - (1-2\alpha)\sin^2 x] \nonumber \\
  & \quad{} \times \exp\left(- \frac{\sqrt{A_6}}{2(1+\alpha)}\sec^2 x\right).  \label{eq:V(1)es} 
\end{align}
As can be checked, the odd wavefunction $\psi_1(x)$ has a single zero at $x=0$ in the defining interval $(-\pi/2, \pi/2)$, since no real $x$ satisfies the condition $\sin^2 x = 3/(1-2\alpha) > 1$.\par
%
%-------------------------------------------------------------------------------------------------
%
\begin{figure}
\begin{center}
\includegraphics{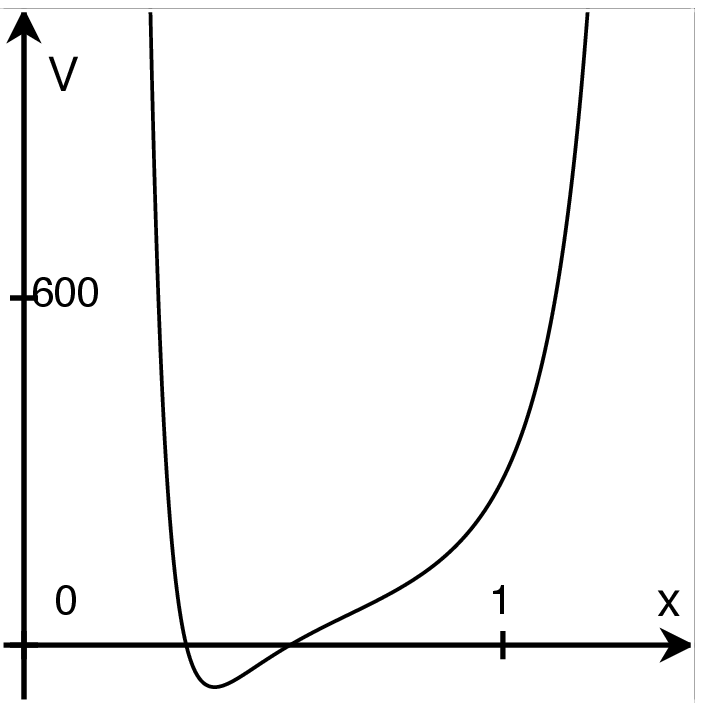}
\caption{Plot of the extended potential $V^{(1,1)}(x)$ with $A_6 = B_6 = 1$ and $\alpha = 1/2$. The ground and first excited state energies are $E_0 = 69/2$ and $E_1 = 293/2$. The PDM reads $m(x) = \left(1 + \frac{1}{2} \cos2x\right)^{-2}$.}
\end{center}
\end{figure}
\par
%
%----------------------------------------------------------------------------------------------------------------
%
\begin{figure}
\begin{center}
\includegraphics{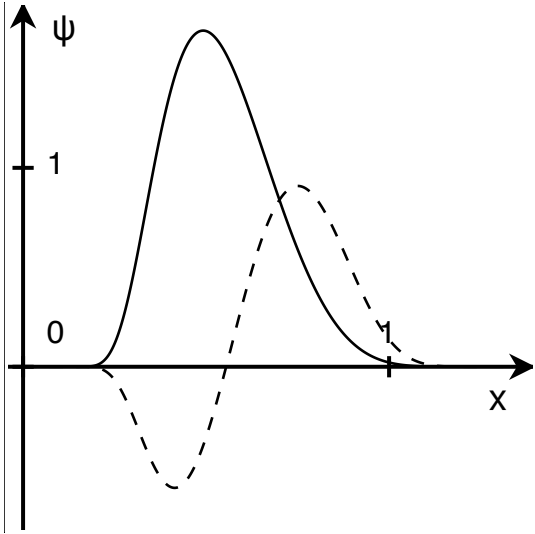}
\caption{Plots of the ground state wavefunction $\psi_0(x)$ (solid line) and of the first excited state wavefunction $\psi_1(x)$ (dashed line) for the potential displayed in fig.~3.}
\end{center}
\end{figure}
\par
%
%-----------------------------------------------------------------------------------------------------------------
%
In fig.~1, an example of extended potential (\ref{eq:V(1)}) is plotted. Its corresponding (rescaled) unnormalized wavefunctions (\ref{eq:V(1)gs}) and (\ref{eq:V(1)es}) are displayed in fig.~2.\par
%
%==========================================================================
%
\section{Extensions of the two-parameter trigonometric P\"oschl-Teller potential}
\setcounter{equation}{0}

Let us consider next an infinite family of extensions of the two-parameter TPT potential (\ref{eq:TPT-2}), defined by
\begin{equation}
  V^{(m_1,m_2)}(x) = \sum_{k=1}^{2m_1+1} A_{2k} \sec^{2k}x + \sum_{l=1}^{2m_2+1} B_{2l} \csc^{2l}x,
  \qquad 0 < x < \frac{\pi}{2},  \label{eq:Vm1m2}
\end{equation}
where $m_1$ and $m_2$ are two nonnegative integers and $A_{4m_1+2}, B_{4m_2+2} > 0$. For $m_1 = m_2 = 0$, eq.~(\ref{eq:Vm1m2}) would give back eq.~(\ref{eq:TPT-2}) with $A_2 = A(A-1)$ and $B_2 = B(B-1)$. Here, we wish to determine parameters $A_2, A_4, \ldots, A_{4m_1}, B_2, B_4, \ldots B_{4m_2}$ in terms of $A_{4m_1+2}$, $B_{4m_2+2}$, and $\alpha$ in such a way that eq. (\ref{eq:def-SE}) with $f(x)$ and $V(x)$ given by (\ref{eq:f-2}) and (\ref{eq:Vm1m2}), respectively, has known ground and first excited states. The corresponding PDM will then be
\begin{equation}
  m(x) = (1 + \alpha \cos2x)^{-2}, \qquad 0 < |\alpha| <1.
\end{equation}
It is enough to assume $m_1 \ge m_2$, because the $m_1<m_2$ case could be easily obtained from that with $m_1>m_2$ by permuting the roles of $\sec^2x$ and $\csc^2x$, which can be achieved by the change of variable $x \to \frac{\pi}{2} - x$ and the change of parameter $\alpha \to - \alpha$. Since the cases $m_1 \ge m_2 > 0$ and $m_1 > m_2=0$ have to be distinguished, we will start with the former, then point out the changes to be made to care for the latter.\par
%
%++++++++++++++++++++++++++++++++++++++++++++++++++++++++++
%
\subsection{\boldmath Extensions with $m_1 \ge m_2 > 0$}

Let us consider the generating functions
\begin{align}
  W_+(x) &= 2\sqrt{A_{4m_1+2}} \sum_{k=0}^{m_1} \binom{m_1+m_2+1}{m_2+k+1} 
      \left(\frac{1+\alpha}{1-\alpha}\right)^{m_1-k} (\tan x)^{2k+1} \nonumber \\
  & \quad{} - 2\sqrt{B_{4m_2+2}} \sum_{l=0}^{m_2} \binom{m_1+m_2+1}{m_1+l+1} 
      \left(\frac{1-\alpha}{1+\alpha}\right)^{m_2-l} (\cot x)^{2l+1}, \label{eq:W_+2}\\
  W_-(x) &= (2m_1+1) (1-\alpha) \tan x - (2m_2+1) (1+\alpha) \cot x.  \label{eq:W_-2}
\end{align}
A calculation similar to that carried out in sec.~4 shows that such functions are compatible and that $E_1-E_0$ is given by
\begin{align}
  E_1-E_0 &= 4 \frac{(m_1+m_2+1)!}{m_1! m_2!} \biggl[\sqrt{A_{4m_1+2}} \frac{(1+\alpha)^{m_1+1}}
       {(1-\alpha)^{m_1}} \nonumber \\
  & \quad{} + \sqrt{B_{4m_2+2}} \frac{(1-\alpha)^{m_2+1}}{(1+\alpha)^{m_2}}\biggr].  \label{eq:diff-2}
\end{align}
\par
%
%----------------------------------------------------------------------------------------------------
%
\begin{figure}
\begin{center}
\includegraphics{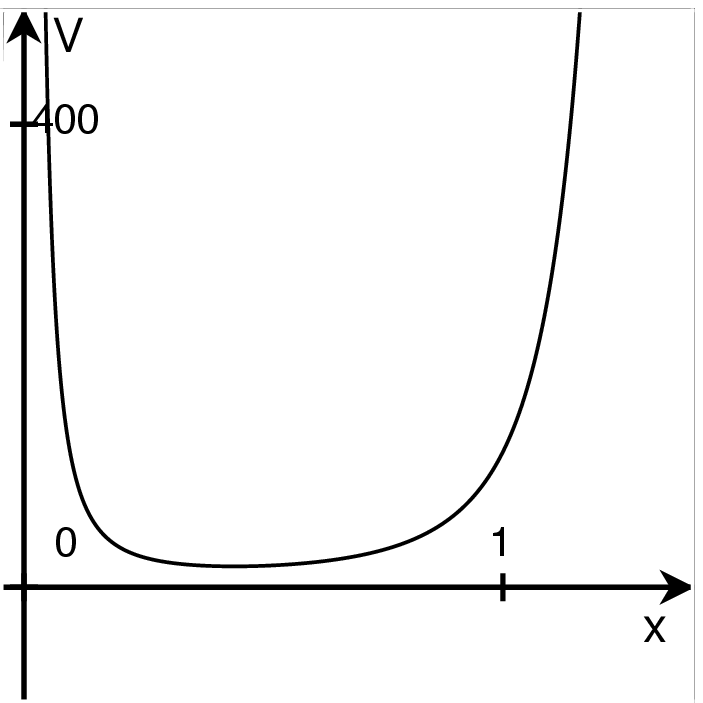}
\caption{Plot of  the extended potential $V^{(1,0)}(x)$ with $A_6 = B_2 = 1$ and $\alpha = 1/2$. The ground and first excited state energies are $E_0 = 629/16$ and $E_1 = 1381/16$. The PDM reads $m(x) = \left(1 + \frac{1}{2} \cos2x\right)^{-2}$.}
\end{center}
\end{figure}
\par
%
%----------------------------------------------------------------------------------------------------------------------
%
\begin{figure}
\begin{center}
\includegraphics{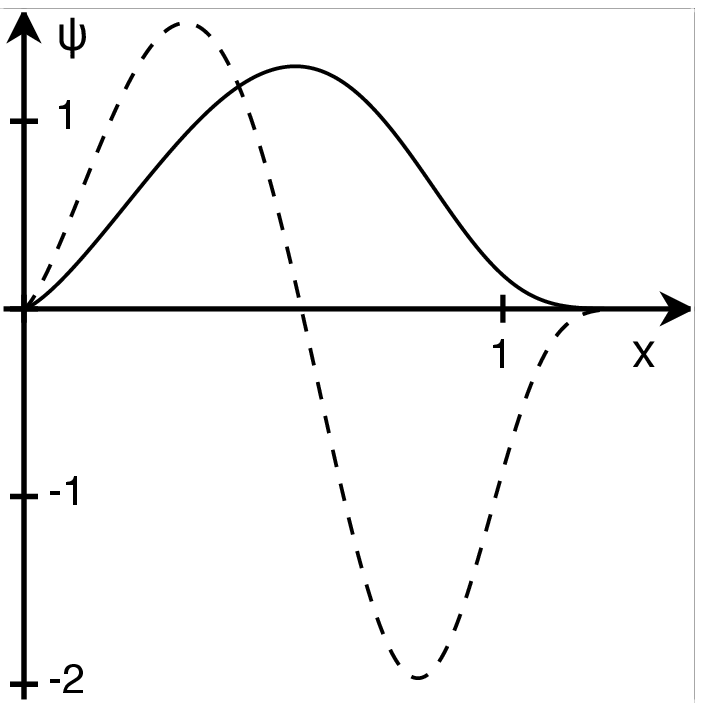}
\caption{Plots of the ground state wavefunction $\psi_0(x)$ (solid line) and of the first excited state wavefunction $\psi_1(x)$ (dashed line) for the potential displayed in fig.~5.}
\end{center}
\end{figure}
\par
%
%-----------------------------------------------------------------------------------------------------------------
%  
The two superpotentials $W(x)$ and $W'(x)$ can now be expressed as
\begin{align}
  W(x) &= \sum_{k=0}^{m_1} \lambda_k (\tan x)^{2k+1} - \sum_{l=0}^{m_2} \mu_l (\cot x)^{2l+1}, \\
  W'(x) &= \sum_{k=0}^{m_1} \lambda'_k (\tan x)^{2k+1} - \sum_{l=0}^{m_2} \mu'_l (\cot x)^{2l+1},
\end{align}
where
\begin{equation}
\begin{split}
  \lambda_0&= \sqrt{A_{4m_1+2}} \binom{m_1+m_2+1}{m_2+1} 
      \left(\frac{1+\alpha}{1-\alpha}\right)^{m_1} - \left(m_1+\frac{1}{2}\right)(1-\alpha), \\
  \lambda'_0 &= \lambda_0 + (2m_1+1)(1-\alpha), \\
  \lambda_k &= \lambda'_k = \sqrt{A_{4m_1+2}} \binom{m_1+m_2+1}{m_2+k+1} 
      \left(\frac{1+\alpha}{1-\alpha}\right)^{m_1-k}, \qquad k=1, 2, \ldots, m_1, \\ 
  \mu_0&= \sqrt{B_{4m_2+2}} \binom{m_1+m_2+1}{m_1+1} 
      \left(\frac{1-\alpha}{1+\alpha}\right)^{m_2} - \left(m_2+\frac{1}{2}\right)(1+\alpha), \\
  \mu'_0 &= \mu_0 + (2m_2+1)(1+\alpha), \\
  \mu_l &= \mu'_l = \sqrt{B_{4m_2+2}} \binom{m_1+m_2+1}{m_1+l+1} 
      \left(\frac{1-\alpha}{1+\alpha}\right)^{m_2-l}, \qquad l=1, 2, \ldots, m_2. 
\end{split}
\end{equation}
\par
%
%--------------------------------------------------------------------------------------------------------
%
As in sec.~4, the determination of $E_0$ and $A_2, A_4, \ldots, A_{4m_1}, B_2, B_4, \ldots, B_{4m_2}$ in terms of $A_{4m_1+2}$, $B_{4m_2+2}$, and $\alpha$ can be carried out in two steps: first to obtain the coefficients $a_k$ and $b_l$ in the expansion
\begin{equation}
  V_1(x) = \sum_{k=0}^{2m_1+1} a_k (\tan x)^{2k} + \sum_{l=1}^{2m_2+1} b_l (\cot x)^{2l},
\end{equation}
then to express the searched for quantities as
\begin{equation}
\begin{split}
  E_0 &= \sum_{k=0}^{2m_1+1} (-1)^{k+1} a_k + \sum_{l=1}^{2m_2+1} (-1)^{l+1} b_l, \\
  A_{2k} &= \sum_{l=k}^{2m_1+1} (-1)^{l-k} \binom{l}{k} a_l, \qquad k=1, 2, \ldots, 2m_1, \\
  B_{2l} &= \sum_{k=l}^{2m_2+1} (-1)^{k-l} \binom{k}{l} b_k, \qquad l=1, 2, \ldots, 2m_2.
\end{split}
\end{equation}
After some lengthy, but straightforward calculations, we get the results detailed in appendix A. \par
%
%----------------------------------------------------------------------------------------------------------
%
To determine the wavefunctions, we rewrite this time $W/f$ and $W'/f$ in terms of the variable $y=\cos 2x$. For the former, for instance, we get
\begin{equation}
  \frac{W(x)}{f(x)} = \sin 2x \left(\sum_{p=1}^{m_1+1} \frac{C_p}{(1+y)^p} - \sum_{q=1}^{m_2+1}
  \frac{D_q}{(1-y)^q} - \frac{\alpha (C_1+D_1)}{1+\alpha y}\right),
\end{equation}
with
\begin{equation}
  C_p = \sum_{q=p-1}^{m_1} 2^q \frac{(-\alpha)^{q-p+1}}{(1-\alpha)^{q-p+2}} \sum_{k=q}^{m_1} (-1)^{k-q}
  \binom{k}{q} \lambda_k, \qquad p=1, 2, \ldots,m_1+1,
\end{equation}
and
\begin{equation}
  D_q = \sum_{p=q-1}^{m_2} 2^p \frac{\alpha^{p-q+1}}{(1+\alpha)^{p-q+2}} \sum_{l=p}^{m_2} (-1)^{l-p}
  \binom{l}{p} \mu_l, \qquad q=1, 2, \ldots, m_2+1.
\end{equation}
The results read
\begin{align}
  \psi_0(x) &\propto f^{- \frac{1}{2}(C_1+D_1+1)} (\cos x)^{C_1} (\sin x)^{D_1} \nonumber \\
  & \quad{} \times \exp\left[- \sum_{p=2}^{m_1+1} \frac{C_p}{2^p (p-1)} (\sec x)^{2(p-1)}
      - \sum_{q=2}^{m_2+1} \frac{D_q}{2^q (q-1)} (\csc x)^{2(q-1)}\right] 
\end{align}
and
\begin{align}
  \psi_1(x) &\propto f^{- \frac{1}{2}(C_1+D_1+2m_1+2m_2+3)} (\cos x)^{C_1} (\sin x)^{D_1} \nonumber \\
  & \quad{} \times \biggl\{-2 \sqrt{B_{4m_2+2}} \sum_{k=0}^{m_2} (-1)^k \binom{m_1+m_2+1}{k}
      \left(\frac{2\alpha}{1+\alpha}\right)^k \sin^{2k}x \nonumber \\
  & \quad {} + \sum_{k=m_2+1}^{m_1+m_2+1} \biggl[\binom{m_1+m_2+1}{k} \nonumber \\
  & \quad{} \times \biggl(2 \sqrt{A_{4m_1+2}} \left(\frac{1+\alpha}{1-\alpha}\right)^{m_1+m_2-k+1}
      F\left(k-m_2-1,k;:\frac{1+\alpha}{1-\alpha}\right) \nonumber \\
  & \quad{} - 2\sqrt{B_{4m_2+2}} (-1)^k F\left(m_2,k; \frac{1-\alpha}{1+\alpha}\right)\biggr)
      \sin^{2k}x\biggr] \biggr\} \nonumber \\
  & \quad{} \times \exp\left[- \sum_{p=2}^{m_1+1} \frac{C_p}{2^p (p-1)} (\sec x)^{2(p-1)}
      - \sum_{q=2}^{m_2+1} \frac{D_q}{2^q (q-1)} (\csc x)^{2(q-1)}\right], 
\end{align}
 where we have defined
 \begin{equation}
  F(n,k;z) = \sum_{p=0}^n (-1)^p \binom{k}{p} z^p, \qquad k>n.
\end{equation}
At the boundaries $x=0$ and $x=\pi/2$ of the defining interval, the behaviour of $\psi_0(x)$ and $\psi_1(x)$ is governed by that of $\exp[-D_{m_2+1} (\csc x)^{2m_2}/(2^{m_2+1}m_2)]$ and $\exp[-C_{m_1+1} (\sec x)^{2m_1}/(2^{m_1+1}m_1)]$, where $D_{m_2+1} = 2^{m_2} \sqrt{B_{4m_2+2}}/(1+\alpha) > 0$ and $C_{m_1+1} = 2^{m_1} \sqrt{A_{4m_1+2}}/(1-\alpha) > 0$, respectively. Hence, such functions are square integrable on $(0,\pi/2)$, as it should be.\par
%
%------------------------------------------------------------------------------------------------------
% 
{}For the simplest case corresponding to $m_1=m_2=1$, we get, for instance, the potential
\begin{align}
  V^{(1,1)}(x) &= \biggl[\frac{15}{4}(1-\alpha)^2 - 24\alpha\sqrt{A_6} + 
      \frac{12\alpha(1+2\alpha)}{(1-\alpha)^2} A_6 - \frac{6(1-\alpha)}{1+\alpha} \sqrt{A_6B_6}\biggr]
      \sec^2 x \nonumber \\
  & \quad{} + 3\sqrt{A_6} \biggl[-2(1-\alpha) + \frac{1+3\alpha}{1-\alpha} \sqrt{A_6}\biggr] \sec^4 x + A_6 
      \sec^6 x \nonumber \\
  & \quad{} + \biggl[\frac{15}{4}(1+\alpha)^2 + 24\alpha\sqrt{B_6}  
      - \frac{6(1+\alpha)}{1-\alpha} \sqrt{A_6B_6} - \frac{12\alpha(1-2\alpha)}{(1+\alpha)^2} B_6 \biggr]
      \csc^2 x \nonumber \\
  & \quad{} + 3\sqrt{B_6} \biggl[-2(1+\alpha) + \frac{1-3\alpha}{1+\alpha} \sqrt{B_6}\biggr] \csc^4 x + B_6 
      \csc^6 x.  \label{eq:V(1,1)}
\end{align}
Its ground and first excited state energies are given by
\begin{align}
  E_0 &= 3(2\alpha^2+3) + \frac{12(\alpha^2-2\alpha-1)}{1-\alpha} \sqrt{A_6} + 
      \frac{12(\alpha^2+2\alpha-1)}{1+\alpha} \sqrt{B_6} \nonumber \\
  & \quad{} + \frac{4(1+2\alpha)^2}{(1-\alpha)^2} A_6 + \frac{8(1-2\alpha)(1+2\alpha)}{(1-\alpha)(1+\alpha)}
      \sqrt{A_6B_6} + \frac{4(1-2\alpha)^2}{(1+\alpha)^2} B_6
\end{align}
and
\begin{align}
  E_1 &= 3(2\alpha^2+3) + \frac{12(3\alpha^2+2\alpha+1)}{1-\alpha} \sqrt{A_6} + 
      \frac{12(3\alpha^2-2\alpha+1)}{1+\alpha} \sqrt{B_6} \nonumber \\
  & \quad{} + \frac{4(1+2\alpha)^2}{(1-\alpha)^2} A_6 + \frac{8(1-2\alpha)(1+2\alpha)}{(1-\alpha)(1+\alpha)}
      \sqrt{A_6B_6} + \frac{4(1-2\alpha)^2}{(1+\alpha)^2} B_6,
\end{align}
with corresponding wavefunctions
\begin{align}
  \psi_0(x) &\propto f^{-\frac{1+\alpha}{(1-\alpha)^2}\sqrt{A_6} -\frac{1-\alpha}{(1+\alpha)^2}\sqrt{B_6}
      +1} (\cos x)^{\frac{2(1+\alpha)}{(1-\alpha)^2}\sqrt{A_6} - \frac{3}{2}} 
      (\sin x)^{\frac{2(1-\alpha)}{(1+\alpha)^2}\sqrt{B_6} - \frac{3}{2}} \nonumber \\
  & \quad{} \times \exp\biggl[- \frac{\sqrt{A_6}}{2(1-\alpha)} \sec^2x - \frac{\sqrt{B_6}}{2(1+\alpha)} 
      \csc^2x\biggr] \label{eq:V(1,1)gs}  
\end{align}
and
\begin{align}
  \psi_1(x) &\propto f^{-\frac{1+\alpha}{(1-\alpha)^2}\sqrt{A_6} -\frac{1-\alpha}{(1+\alpha)^2}\sqrt{B_6}
      -2} (\cos x)^{\frac{2(1+\alpha)}{(1-\alpha)^2}\sqrt{A_6} - \frac{3}{2}} 
      (\sin x)^{\frac{2(1-\alpha)}{(1+\alpha)^2}\sqrt{B_6} - \frac{3}{2}} \nonumber \\
  & \quad{} \times \biggl\{- 2\sqrt{B_6} + \frac{12\alpha}{1+\alpha} \sqrt{B_6} \sin^2x + 6 \biggl[
      \frac{1+\alpha}{1-\alpha} \sqrt{A_6} + \frac{1-3\alpha}{1+\alpha} \sqrt{B_6}\biggr] \sin^4x \nonumber \\
  & \quad{} - 4\biggl[\frac{1+2\alpha}{1-\alpha} \sqrt{A_6} + \frac{1-2\alpha}{1+\alpha} \sqrt{B_6}\biggr]
      \sin^6x\biggr\} \nonumber \\
  & \quad{} \times \exp\biggl[- \frac{\sqrt{A_6}}{2(1-\alpha)} \sec^2x - \frac{\sqrt{B_6}}{2(1+\alpha)} 
      \csc^2x\biggr].  \label{eq:V(1,1)es} 
\end{align}
\par
%
%-----------------------------------------------------------------------------------------------------
%
In fig.~3, an example of extended potential (\ref{eq:V(1,1)}) is plotted. Its corresponding (rescaled) unnormalized wavefunctions (\ref{eq:V(1,1)gs}) and (\ref{eq:V(1,1)es}) are displayed in fig.~4.\par
%
%++++++++++++++++++++++++++++++++++++++++++++++++++++++++++++++++
% 
\subsection{\boldmath Extensions with $m_1> m_2=0$}

{}For $m_1> m_2=0$, the results presented in sec.~5.1 remain valid provided we replace $\sqrt{B_2}$ by $1 + \alpha + \frac{1}{2}\Delta$, $\Delta = \sqrt{(1+\alpha)^2 + 4B_2}$, in the generating functions (\ref{eq:W_+2}) and (\ref{eq:W_-2}), which therefore become
\begin{align}
  W_+(x) &= 2\sqrt{A_{4m_1+2}} \sum_{k=0}^{m_1} \binom{m_1+1}{k+1} 
     \left(\frac{1+\alpha}{1-\alpha}\right)^{m_1-k} (\tan x)^{2k+1} \nonumber \\
  & \quad{} - (2+2\alpha+\Delta) \cot x, \\
  W_-(x) &= (2m_1+1)(1-\alpha) \tan x - (1+\alpha) \cot x,
\end{align}
with corresponding $E_1-E_0$ given by
\begin{equation}
  E_1 - E_0 = 4(m_1+1) \frac{(1+\alpha)^{m_1+1}}{(1-\alpha)^{m_1}} \sqrt{A_{4m_1+2}} 
  + 2(m_1+1)(1-\alpha)(2+2\alpha+\Delta).
\end{equation}
\par
%
%----------------------------------------------------------------------------------------------------------
%
We shall not present the general results, but instead show the simplest example corresponding to $m_1=1$. In such a case, the potential reads
\begin{align}
  V^{(1,0)}(x) &= \biggl[\frac{15}{4}(1-\alpha)^2 - (24\alpha+\Delta) \sqrt{A_6} + 
      \frac{-1+2\alpha+15\alpha^2}{(1-\alpha)^2} A_6\biggr] \sec^2x \nonumber \\
  & \quad{} + \sqrt{A_6} \biggl[-6(1-\alpha) + \frac{1+7\alpha}{1-\alpha}\sqrt{A_6}\biggr] \sec^4x
      + A_6 \sec^6x + B_2 \csc^2x,  \label{eq:V(1,0)}
\end{align}
with ground and first excited state energies
\begin{align}
  E_0 &= \frac{5}{4}(1-\alpha)(1-5\alpha) - (1-\alpha)\Delta + 
     \frac{-2-4\alpha+22\alpha^2+(1+3\alpha)\Delta}{1-\alpha} \sqrt{A_6} \nonumber \\
  & \quad{} + \left(\frac{1+3\alpha}{1-\alpha}\right)^2 A_6 + B_2, \\
  E_1 &= \frac{1}{4}(1-\alpha)(37+7\alpha) + 3(1-\alpha)\Delta + 
     \frac{6(1+2\alpha+5\alpha^2)+(1+3\alpha)\Delta}{1-\alpha} \sqrt{A_6} \nonumber \\
  & \quad{} + \left(\frac{1+3\alpha}{1-\alpha}\right)^2 A_6 + B_2,  
\end{align}
and corresponding wavefunctions
\begin{align}
  \psi_0(x) &\propto f^{-\frac{1+\alpha}{2(1-\alpha)^2}\sqrt{A_6} - \frac{\Delta}{4(1+\alpha)}}
      (\cos x)^{\frac{1+\alpha}{(1-\alpha)^2}\sqrt{A_6} - \frac{3}{2}} (\sin x)^{\frac{\Delta}{2(1+\alpha)}
      +\frac{1}{2}} \nonumber \\
  & \quad{} \times \exp\left[- \frac{\sqrt{A_6}}{2(1-\alpha)} \sec^2x\right],  \label{eq:V(1,0)gs} \\
  \psi_1(x) &\propto f^{-\frac{1+\alpha}{2(1-\alpha)^2}\sqrt{A_6} - \frac{\Delta}{4(1+\alpha)} - 2}
      (\cos x)^{\frac{1+\alpha}{(1-\alpha)^2}\sqrt{A_6} - \frac{3}{2}} (\sin x)^{\frac{\Delta}{2(1+\alpha)}
      +\frac{1}{2}} \nonumber \\
  & \quad{} \times \biggl\{2+2\alpha+\Delta - 2\biggl[\frac{2(1+\alpha)}{1-\alpha}\sqrt{A_6} +2+2\alpha
      +\Delta\biggr] \sin^2x \nonumber \\
  & \quad{} + \biggl[\frac{2(1+3\alpha)}{1-\alpha}\sqrt{A_6} +2+2\alpha+\Delta\biggr] \sin^4x\biggr\}
      \exp\left[- \frac{\sqrt{A_6}}{2(1-\alpha)} \sec^2x\right].  \label{eq:V(1,0)es}
\end{align}
\par
%
%-------------------------------------------------------------------------------------------------
%
In fig.~5, an example of extended potential (\ref{eq:V(1,0)}) is plotted. Its corresponding (rescaled) unnormalized wavefunctions (\ref{eq:V(1,0)gs}) and (\ref{eq:V(1,0)es}) are displayed in fig.~6.\par
%
%============================================================
%
\section{Conclusion}

In the present paper, we have shown that it is possible to generate infinite families of PDM Schr\"odinger equations with known ground and first excited states in DSUSY by considering extensions of both one- and two-parameter TPT potentials endowed with a DSI property. If needed, higher energy levels should be calculated numerically. This work completes a previous study \cite{cq18}, where only extensions of some simpler potentials were explicitly constructed, and demonstrates the efficiency of the method proposed there to deal with more complex potentials. This opens the way for building extensions of other potentials with a DSI property, whose treatment is rather involved, such as the Eckart and Rosen-Morse I potentials considered in \cite{bagchi05, cq09}.\par
%
%------------------------------------------------------------------------------------------
%
Taking into account the usefulness of the TPT potential as a first approximation in several problems of molecular and solid state physics, it is obvious that the exact results presented here for potentials including some extra terms may find helpful applications in such fields. The search for such applications would be another interesting topic for future investigation.\par
%
%==============================================================
%
\section*{\boldmath Appendix A. General results for extensions of the two-parameter trigonometric P\"oschl-Teller potential with $m_1 \ge m_2 > 0$}

\renewcommand{\theequation}{A.\arabic{equation}}
\setcounter{equation}{0}

In this appendix, we present the general results obtained for the ground state energy and the parameters of the extensions of the two-parameter trigonometric P\"oschl-Teller potential with $m_1 \ge m_2 > 0$:
\begin{align}
  E_0 &= (m_1+m_2+1)^2 - 2\alpha(m_1-m_2)(m_1+m_2+2) + \alpha^2[(m_1-m_2)^2 \nonumber \\
  &\quad{} + 2(m_1+m_2+1)] \nonumber \\
  &\quad{} - \sqrt{A_{4m_1+2}} \biggl\{2m_2 \binom{m_1+m_2+1}{m_2+1} \frac{(1+\alpha)^{m_1+1}}
      {(1-\alpha)^{m_1}} \nonumber \\
  &\quad{} + \sum_{k=1}^{m_1+1} (-1)^k \biggl[(2m_2-2k) \binom{m_1+m_2+1}{m_2+k+1} 
       - (2m_1+2k) \binom{m_1+m_2+1}{m_2+k}\biggr] \nonumber \\
  &\quad{} \times \frac{(1+\alpha)^{m_1-k+1}}{(1-\alpha)^{m_1-k}}\Biggr\} \nonumber \\
  &\quad{} - \sqrt{B_{4m_2+2}} \biggl\{2m_1 \binom{m_1+m_2+1}{m_1+1} \frac{(1-\alpha)^{m_2+1}}
      {(1+\alpha)^{m_2}} \nonumber \\
  &\quad{} + \sum_{l=1}^{m_2+1} (-1)^l \biggl[(2m_1-2l) \binom{m_1+m_2+1}{m_1+l+1} 
       - (2m_2+2l) \binom{m_1+m_2+1}{m_1+l}\biggr] \nonumber \\
  &\quad{} \times \frac{(1-\alpha)^{m_2-l+1}}{(1+\alpha)^{m_2-l}}\Biggr\} \nonumber \\
  &\quad{} - A_{4m_1+2} \sum_{k=1}^{2m_1+1} \biggl[(-1)^k \left(\frac{1+\alpha}
      {1-\alpha}\right)^{2m_1-k+1} \nonumber \\
  & \quad{} \times \sum_{l={\rm max}(0,k-m_1-1)}^{{\rm min}(k-1,m_1)} \binom{m_1+m_2+1}{m_2+l+1}
      \binom{m_1+m_2+1}{m_2+k-l}\biggr] \nonumber \\
  & \quad{} + 2\sqrt{A_{4m_1+2}B_{4m_2+2}} \biggl\{\sum_{k=0}^{m_1} \biggl[(-1)^k 
      \left(\frac{1+\alpha}{1-\alpha}\right)^{m_1-m_2-k} \nonumber \\
  & \quad{} \times \sum_{l=k}^{{\rm min}(m_2+k,m_1)} \binom{m_1+m_2+1}{m_2+l+1}
      \binom{m_1+m_2+1}{m_2+k-l}\biggr] \nonumber \\
  & \quad{} + \sum_{l=1}^{m_2} \biggl[(-1)^l 
      \left(\frac{1+\alpha}{1-\alpha}\right)^{m_1-m_2+l}\sum_{k=l}^{m_2} \binom{m_1+m_2+1}{m_1+k+1}
      \binom{m_1+m_2+1}{m_1+l-k}\biggr]\biggr\} \nonumber \\
  &\quad{} - B_{4m_2+2} \sum_{l=1}^{2m_2+1} \biggl[(-1)^l \left(\frac{1-\alpha}
      {1+\alpha}\right)^{2m_2-l+1} \nonumber \\
  & \quad{} \times \sum_{k={\rm max}(0,l-m_2-1)}^{{\rm min}(l-1,m_2)} \binom{m_1+m_2+1}{m_1+k+1}
      \binom{m_1+m_2+1}{m_1+l-k}\biggr],  \label{eq:E0-2}
\end{align}
\begin{align}
  A_2 &= \left(m_1+\frac{1}{2}\right) \left(m_1+\frac{3}{2}\right) (1-\alpha)^2 \nonumber \\
  & \quad{} - \sqrt{A_{4m_1+2}} \sum_{k=1}^{m_1+1} \biggl\{(-1)^k k \frac{(1+\alpha)^{m_1-k+1}}
      {(1-\alpha)^{m_1-k}} \nonumber \\
  & \quad{} \times \biggl[(2m_2-2k) \binom{m_1+m_2+1}{m_2+k+1} - (2m_1+2k) 
      \binom{m_1+m_2+1}{m_2+k}\biggr]\biggr\} \nonumber \\
  &\quad{} - A_{4m_1+2} \sum_{k=1}^{2m_1+1} \biggl[(-1)^k k \left(\frac{1+\alpha}
      {1-\alpha}\right)^{2m_1-k+1} \nonumber \\
  & \quad{} \times \sum_{l={\rm max}(0,k-m_1-1)}^{{\rm min}(k-1,m_1)} \binom{m_1+m_2+1}{m_2+l+1}
      \binom{m_1+m_2+1}{m_2+k-l}\biggr] \nonumber \\
  & \quad{} + 2\sqrt{A_{4m_1+2}B_{4m_2+2}} \sum_{k=1}^{m_1} \biggl[(-1)^k k 
      \left(\frac{1+\alpha}{1-\alpha}\right)^{m_1-m_2-k} \nonumber \\
  & \quad{} \times \sum_{l=k}^{{\rm min}(m_2+k,m_1)} \binom{m_1+m_2+1}{m_2+l+1}
      \binom{m_1+m_2+1}{m_2+k-l}\biggr],
\end{align}
\begin{align}
  A_{2k} &= \sqrt{A_{4m_1+2}} \sum_{l=k}^{m_1+1} \biggl\{(-1)^{l-k} \binom{l}{k} 
       \frac{(1+\alpha)^{m_1-l+1}}{(1-\alpha)^{m_1-l}} \nonumber \\
  & \quad{} \times \biggl[(2m_2-2l) \binom{m_1+m_2+1}{m_2+l+1} - (2m_1+2l) 
      \binom{m_1+m_2+1}{m_2+l}\biggr]\biggr\} \nonumber \\
  & \quad{} + A_{4m_1+2} \sum_{l=k}^{2m_1+1} \biggl[(-1)^{l-k} \binom{l}{k} \left(\frac{1+\alpha}
      {1-\alpha}\right)^{2m_1-l+1} \nonumber \\
  & \quad{} \times \sum_{p={\rm max}(0,l-m_1-1)}^{{\rm min}(l-1,m_1)} \binom{m_1+m_2+1}{m_2+p+1}
      \binom{m_1+m_2+1}{m_2+l-p}\biggr] \nonumber \\
  & \quad{} - 2\sqrt{A_{4m_1+2}B_{4m_2+2}} \sum_{l=k}^{m_1} \biggl[(-1)^{l-k} \binom{l}{k} 
      \left(\frac{1+\alpha}{1-\alpha}\right)^{m_1-m_2-l} \nonumber \\
  & \quad{} \times \sum_{p=l}^{{\rm min}(m_2+l,m_1)} \binom{m_1+m_2+1}{m_2+p+1}
      \binom{m_1+m_2+1}{m_2+l-p}\biggr],  \qquad 2 \le k \le m_1+1, 
\end{align}
\begin{align}
  A_{2k} &= A_{4m_1+2} \sum_{l=k}^{2m_1+1} \biggl[(-1)^{l-k} \binom{l}{k} \left(\frac{1+\alpha}
      {1-\alpha}\right)^{2m_1-l+1} \nonumber \\
  & \quad{} \times \sum_{p=l-m_1-1}^{m_1} \binom{m_1+m_2+1}{m_2+p+1}
      \binom{m_1+m_2+1}{m_2+l-p}\biggr], \qquad  m_1+2 \le k \le 2m_1,
\end{align}
\begin{align}
  B_2 &= \left(m_2+\frac{1}{2}\right) \left(m_2+\frac{3}{2}\right) (1+\alpha)^2 \nonumber \\
  & \quad{} - \sqrt{B_{4m_2+2}} \sum_{k=1}^{m_2+1} \biggl\{(-1)^k k \frac{(1-\alpha)^{m_2-k+1}}
      {(1+\alpha)^{m_2-k}} \nonumber \\
  & \quad{} \times \biggl[(2m_1-2k) \binom{m_1+m_2+1}{m_1+k+1} - (2m_2+2k) 
      \binom{m_1+m_2+1}{m_1+k}\biggr]\biggr\} \nonumber \\
  & \quad{} + 2\sqrt{A_{4m_1+2}B_{4m_2+2}} \sum_{k=1}^{m_2} \biggl[(-1)^k k 
      \left(\frac{1+\alpha}{1-\alpha}\right)^{m_1-m_2+k} \nonumber \\
  & \quad{} \times \sum_{l=k}^{m_2} \binom{m_1+m_2+1}{m_1+l+1}
      \binom{m_1+m_2+1}{m_1+k-l}\biggr] \nonumber \\
  &\quad{} - B_{4m_2+2} \sum_{k=1}^{2m_2+1} \biggl[(-1)^k k \left(\frac{1-\alpha}
      {1+\alpha}\right)^{2m_2-k+1} \nonumber \\
  & \quad{} \times \sum_{l={\rm max}(0,k-m_2-1)}^{{\rm min}(k-1,m_2)} \binom{m_1+m_2+1}{m_1+l+1}
      \binom{m_1+m_2+1}{m_1+k-l}\biggr],
\end{align}
\begin{align}
  B_{2l} &= \sqrt{B_{4m_2+2}} \sum_{k=l}^{m_2+1} \biggl\{(-1)^{k-l} \binom{k}{l} 
       \frac{(1-\alpha)^{m_2-k+1}}{(1+\alpha)^{m_2-k}} \nonumber \\
  & \quad{} \times \biggl[(2m_1-2k) \binom{m_1+m_2+1}{m_1+k+1} - (2m_2+2k) 
      \binom{m_1+m_2+1}{m_1+k}\biggr]\biggr\} \nonumber \\
  & \quad{} - 2\sqrt{A_{4m_1+2}B_{4m_2+2}} \sum_{k=l}^{m_2} \biggl[(-1)^{k-l} \binom{k}{l} 
      \left(\frac{1+\alpha}{1-\alpha}\right)^{m_1-m_2+k} \nonumber \\
  & \quad{} \times \sum_{p=k}^{m_2} \binom{m_1+m_2+1}{m_1+p+1}
      \binom{m_1+m_2+1}{m_1+k-p}\biggr] \nonumber \\
  & \quad{} + B_{4m_2+2} \sum_{k=l}^{2m_2+1} \biggl[(-1)^{k-l} \binom{k}{l} \left(\frac{1-\alpha}
      {1+\alpha}\right)^{2m_2-k+1} \nonumber \\
  & \quad{} \times \sum_{p={\rm max}(0,k-m_2-1)}^{{\rm min}(k-1,m_2)} \binom{m_1+m_2+1}{m_1+p+1}
      \binom{m_1+m_2+1}{m_1+k-p}\biggr],  \quad 2 \le l \le m_2+1, 
\end{align}
\begin{align}
  B_{2l} &= B_{4m_2+2} \sum_{k=l}^{2m_2+1} \biggl[(-1)^{k-l} \binom{k}{l} \left(\frac{1-\alpha}
      {1+\alpha}\right)^{2m_2-k+1} \nonumber \\
  & \quad{} \times \sum_{p=k-m_2-1}^{m_2} \binom{m_1+m_2+1}{m_1+p+1}
      \binom{m_1+m_2+1}{m_1+k-p}\biggr], \qquad  m_2+2 \le l \le 2m_2.
\end{align}
Note that $E_1$ can be easily obtained from (\ref{eq:diff-2}) and (\ref{eq:E0-2}).\par
% 
%=============================================================
% 

%
%===================================================================
%


\newpage

\begin{thebibliography}{99}

\bibitem{poschl}
G.\ P\"oschl, E.\ Teller, 
Z.\ Phys.\ {\bf 83}, 143 (1933).

\bibitem{flugge}
S.\ Fl\"ugge,
{\sl Practical Quantum Mechanics I} (Springer-Verlag, Berlin, 1971).

\bibitem{antoine}
J.-P.\ Antoine, J.-P.\ Gazeau, P.\ Monceau, J.R.\ Klauder, K.A.\ Penson,
J.\ Math.\ Phys.\ {\bf 42}, 2349 (2001).

\bibitem{scarf}
F.L.\ Scarf,
Phys.\ Rev.\ {\bf 112}, 1137 (1958).

\bibitem{cq12}
C.\ Quesne,
J.\ Phys.\ Conf.\ Ser.\ {\bf 380}, 012016 (2012). 

\bibitem{genden}
L.E.\ Gendenshtein,
JETP Lett.\ {\bf 38}, 356 (1983).

\bibitem{cooper}
F.\ Cooper, A.\ Khare, U.\ Sukhatme,
Phys.\ Rep.\ {\bf 251}, 267 (1995).

\bibitem{contreras}
A.\ Contreras-Astorga, D.J.\ Fern\'andez C.,
J.\ Phys.\ A {\bf 41}, 475303 (2008).

\bibitem{cq08}
C.\ Quesne,
J.\ Phys.\ A {\bf 41}, 392001 (2008).

\bibitem{odake09}
S.\ Odake, R.\ Sasaki,
Phys.\ Lett.\ B {\bf 679}, 414 (2009).

\bibitem{odake11}
S.\ Odake, R.\ Sasaki,
Phys.\ Lett.\ B {\bf 702}, 164 (2011).

\bibitem{gomez14}
D.\ G\'omez-Ullate, Y.\ Grandati, R.\ Milson,
J.\ Math.\ Phys.\ {\bf 55}, 043510 (2014).

\bibitem{bagchi15}
B.\ Bagchi, Y.\ Grandati, C.\ Quesne,
J.\ Math.\ Phys.\ {\bf 56}, 062103 (2015).

\bibitem{grandati}
Y.\ Grandati, C.\ Quesne,
SIGMA {\bf 11}, 061 (2015).

\bibitem{gomez09}
D.\ G\'omez-Ullate, N.\ Kamran, R.\ Milson,
J.\ Math.\ Anal.\ Appl.\ {\bf 359}, 352 (2009).

\bibitem{calogero}
F.\ Calogero, G.\ Yi,
J.\ Phys.\ A {\bf 45}, 095206 (2012).

\bibitem{fernandez}
D.J.\ Fern\'andez C., E.\ Salinas-Hern\'andez, J.\ Phys.\ A {\bf 36}, 2537 (2003). 

\bibitem{bastard}
G.\ Bastard,
{\sl Wave Mechanics Applied to Semiconductor Heterostructures} (Editions de Physique, Les Ulis, 1988).

\bibitem{weisbuch}
C.\ Weisbuch, B.\ Vinter,
{\sl Quantum Semiconductor Heterostructures} (Academic, New York, 1997).

\bibitem{serra}
L.\ Serra, E.\ Lipparini,
Europhys.\ Lett.\ {\bf 40}, 667 (1997).

\bibitem{harrison}
P.\ Harrison, A.\ Valavanis,
{\sl Quantum Wells, Wires and Dots: Theoretical and Computational Physics of Semiconductor Nanostructures} (Wiley, Chichester, 2016).

\bibitem{barranco}
M.\ Barranco, M.\ Pi, S.M.\ Gatica, E.S.\ Hern\'andez, J.\ Navarro, 
Phys.\ Rev.\ B {\bf 56}, 8997 (1997).

\bibitem{geller}
M.R.\ Geller, W.\ Kohn, 
Phys.\ Rev.\ Lett.\ {\bf 70}, 3103 (1993).

\bibitem{arias}
F.\ Arias de Saavedra, J.\ Boronat, A.\ Polls, A.\ Fabrocini, 
Phys.\ Rev.\ B {\bf 50}, 4248 (1994).

\bibitem{puente}
A.\ Puente, Ll.\ Serra, M.\ Casas,
Z.\ Phys.\ D {\bf 31}, 283 (1994).

\bibitem{ring}
P.\ Ring, P.\ Schuck,
{\sl The Nuclear Many Body Problem} (Springer, New York, 1980).

\bibitem{bonatsos}
D.\ Bonatsos, P.E.\ Georgoudis, D.\ Lenis, N.\ Minkov, C.\ Quesne,
Phys.\ Rev.\ C {\bf 83}, 044321 (2011).

\bibitem{willatzen}
W.\ Willatzen, B.\ Lassen,
J.\ Phys.:\ Condens.\ Matter {\bf 19}, 136217 (2007).

\bibitem{chamel}
N.\ Chamel,
Nucl.\ Phys.\ A {\bf 773}, 263 (2006).

\bibitem{cq06}
C.\ Quesne,
Ann.\ Phys.\ (NY) {\bf 321}, 1221 (2006).

\bibitem{cq04}
C.\ Quesne, V.M.\ Tkachuk,
J.\ Phys.\ A {\bf 37}, 4267 (2004).

\bibitem{bagchi05}
B.\ Bagchi, A.\ Banerjee, C.\ Quesne, V.M.\ Tkachuk,
J.\ Phys.\ A {\bf 38}, 2929 (2005).

\bibitem{cq09}
C.\ Quesne,
SIGMA {\bf 5}, 046 (2009).

\bibitem{cq18}
C.\ Quesne,
Ann.\ Phys.\ (NY) {\bf 399}, 270 (2018).

\bibitem{voznyak}
O.\ Voznyak, V.M.\ Tkachuk,
J.\ Phys.\ Stud.\ {\bf 16}, 1003 (2012) (in Ukrainian).

\bibitem{tkachuk}
V.M.\ Tkachuk,
Phys.\ Lett.\ A {\bf 245}, 177 (1998).

\bibitem{vonroos}
O.\ von Roos,
Phys.\ Rev.\ B {\bf 27}, 7547 (1983).

\bibitem{mustafa}
O.\ Mustafa, S.H.\ Mazharimousavi,
Int.\ J.\ Theor.\ Phys.\ {\bf 46}, 1786 (2007).

\end{thebibliography}
\end{document}